# Lowest-ID with Adaptive ID Reassignment: A Novel Mobile Ad-Hoc Networks Clustering Algorithm

Damianos Gavalas[1], Grammati Pantziou[2], Charalampos Konstantopoulos[3], Basilis Mamalis[2]

[1] Department of Cultural Technology and Communication, University of the Aegean, Greece, dgavalas@aegean.gr
[2] Department of Informatics, Technological Education Institute of Athens, Greece, {pantziou, vmamalis}@teiath.gr
[3] Computer Technology Institute, Patras, Greece, konstant@cti.gr

*Abstract* - **Clustering is a promising approach for building hierarchies and simplifying the routing process in mobile ad-hoc network environments. The main objective of clustering is to identify suitable node representatives, i.e. cluster heads (CHs), to store routing and topology information and maximize clusters stability. Traditional clustering algorithms suggest CH election exclusively based on node IDs or location information and involve frequent broadcasting of control packets, even when network topology remains unchanged. More recent works take into account additional metrics (such as energy and mobility) and optimize initial clustering. However, in many situations (e.g. in relatively static topologies) re-clustering procedure is hardly ever invoked; hence initially elected CHs soon reach battery exhaustion. Herein, we introduce an efficient distributed clustering algorithm that uses both mobility and energy metrics to provide stable cluster formations. CHs are initially elected based on the time and cost-efficient lowest-ID method. During clustering maintenance phase though, node IDs are re-assigned according to nodes mobility and energy status, ensuring that nodes with low-mobility and sufficient energy supply are assigned low IDs and, hence, are elected as CHs. Our algorithm also reduces control traffic volume since broadcast period is adjusted according to nodes mobility pattern: we employ infrequent broadcasting for relative static network topologies, and increase broadcast frequency for highly mobile network configurations. Simulation results verify that energy consumption is uniformly distributed among network nodes and that signaling overhead is significantly decreased.**

## I. INTRODUCTION

Hierarchical organization of networks is a well-known and studied problem of distributed computing. It has been proved an effective solution for problems such as, minimizing the amount of storage communication (e.g. routing and multicast tables), thus reducing information update overhead, optimizing the use of network bandwidth and distributed resources throughout the network, etc [1]. While the hierarchical organization fits well in wired infrastructured networks, its suitability in Mobile Ad-Hoc Networks (MANETs) remains an open research issue.

MANETs represent dynamic wireless environments that have been intensively researched within the last years. Unlike wireless cellular networks which rely on a wired backbone connecting base stations, MANETs are self-organizing and self-configuring multi-hop networks where the network structure changes dynamically due to node mobility [4]. MANETs are expected to play a critical role in cases where a wired (central) backbone is neither available nor economical to build, such as law enforcement operations, battle field communications, disaster recovery situations, and so on [11]. Such situations require a dynamic network topology where all nodes, including routers, are mobile and communication between two end nodes may be supported by intermediate nodes.

Similarly to wired networks, flat MANET structures encounter scalability problems with increased network size, especially in the face of node mobility due to MANETs intrinsic characteristics. Hence, the need for partitioning MANET nodes among virtual groups is imperative. Virtual grouping would create hierarchies of nodes, such that the network topology can be abstracted. This process is commonly referred to as *clustering* and the substructures that are collapsed in higher levels are called *clusters* [3]. Clustering is also crucial for controlling the spatial reuse of the shared channel (e.g. in terms of time division and frequency division schemes), for minimizing the amount of data to be exchanged in order to maintain routing and control information in a mobile environment, as well as for building and maintaining cluster-based virtual network architectures.

In clustering procedure, a representative of each cluster is 'elected' as a *cluster head* (CH) and a node that belongs to more than two clusters is called *gateway*. Remaining members are called *ordinary nodes*. CHs hold routing and topology information, relaxing ordinary mobile hosts (MHs) from such requirement; however, they represent network bottleneck points and -being engaged in packet forwarding activities- are prone to fast battery exhaustion. The boundaries of a cluster are defined by the transmission area of its CH.

The concept of clustering in MANETs is not new; many algorithms that consider different metrics and focus on diverse objectives have been proposed [2][6][7][9]. Most existing schemes separate clustering into two phases, cluster formation and cluster maintenance (where initial cluster configurations may be modified, depending on nodes movement). During cluster maintenance phase those algorithms typically involve increased exchange of control messages and fail to preserve valuable energy resources of CHs. In this article, we introduce a distributed algorithm for efficient and scalable clustering of MANETs that corrects the two aforementioned problems. The main contributions of the algorithm are: fast and inexpensive

The research work presented herein has been co-funded by 75% from EU and 25% from the Greek government under the framework of the Education and Initial Vocational Training II, Programme Archimedes.

completion of clustering procedure; incorporation of both mobility and battery power metrics in cluster formation; fairness in cumulative time of serving as CHs among network nodes; minimization of control traffic volume during clustering maintenance phase.

The remainder of the paper is organized as follows: Section II provides an overview of related work in the field of cluster-based mobile ad-hoc networks. Section III describes the details of our proposed algorithm, while Section IV discusses simulation results. Finally, Section V concludes the paper and draws directions for future work.

## II. RELATED WORK

Several heuristics have been proposed to address ad-hoc networks clustering problem. One of the most popular ones is the Lowest-ID (LID) [9], wherein each node is assigned a unique ID. Periodically, nodes broadcast their IDs through a 'Hello' control message, within a period termed 'Hello period' (HP). The lowest-ID node in a neighborhood is then elected as the CH; nodes which can 'hear' two or more CHs become gateways, while remaining MHs are considered as ordinary nodes.

Highest-Degree (HD) algorithm, originally proposed in [7], exclusively uses location information for cluster formation: the highest *degree* node in a neighborhood, i.e. the node with the largest number of neighbors is elected as CH. Experiments demonstrate that the system is not scalable: as the number of nodes in a cluster is increased, the throughput drops and hence a gradual degradation in the system performance is observed. Moreover, in highly mobile environments, the re-affiliation rate increases due to node movements and as a result, the highest-degree node (the current CH) may fail to be re-elected even if it looses a single neighbor [2].

The main asset of LID method is its implementation simplicity. It is also a quick clustering method, as it only takes two HPs to decide upon cluster structure and also provides a more stable cluster formation than HD. In contrast, HD needs three HPs to establish a clustered architecture [6]. However, the main drawback of LID heuristic is its bias towards nodes with smaller IDs: these nodes are highly likely to serve as CHs for long periods which may lead to their rapid battery drainage. In addition, neither LID nor HD algorithm take into account mobility metrics, i.e. highly mobile nodes are equally likely to be elected as CHs, although their movement away from their attached cluster members may soon lead to a ripple re-clustering effect [12].

The Weighted Clustering Algorithm (WCA) [2] employs combined-metrics-based clustering: a number of metrics, including node degree, CH serving time (to estimate residual energy capacity) and moving speed, are taken into account to calculate a weight factor $I_v$ for every node *v*. Mobile nodes with local minimum $I_v$ are elected as CHs. CHs election process is invoked at the very beginning of cluster formation or when a mobile node moves to a region not covered by any CH. WCA does not invoke re-clustering when a member node changes its attaching cluster. Even though this mechanism can enhance the stability of cluster topology, this also implies that CHs keep their status without considering the attribute of minimum $I_v$ in later cluster maintenance. For instance, in relatively static networking environments, WCA will hardly ever be invoked, hence CHs service time will be prolonged and elected CHs will soon suffer from battery exhaustion. Also, the CH serving time alone is not a reliable indicator of energy consumption; hence, the accuracy of $I_v$ values in WCA execution is in doubt [12].

## III. LOWEST-ID WITH ADAPTIVE ID REASSIGNMENT (LIDAR) ALGORITHM

Mobility is a prominent characteristic of MANETs, and is the main factor affecting topology change and route invalidation [2]. Mobile nodes that exhibit high mobility are inadequate for serving as CHs since their movement is likely to trigger frequent re-clustering, therefore increasing control traffic volume.

In addition, mobile nodes in a MANET normally depend on battery power supply during operation, hence the energy limitation poses a severe challenge for network performance [11][12]. A MANET should strive to reduce its energy consumption greedily in order to prolong the network lifespan. Also, a CH bears extra work compared with ordinary members, and it more likely "dies" early because of excessive energy consumption. The lack of mobile nodes due to energy depletion may cause network partition and communication interruption [11]. Hence, it is also important to balance the energy consumption among mobile nodes to avoid node failures.

In this article, we propose a novel clustering algorithm, Lowest-ID with Adaptive ID Reassignment (LIDAR). The main idea behind LIDAR is to maintain the assets of LID algorithm (fast, simple and low-cost clustering process) while providing stable clusters and catering for balanced computational load and power consumption among mobile nodes (to maximize scalability and extend the network's lifespan). This is achieved by identifying and electing the most suitable nodes as CHs, i.e. those with sufficient power level and low mobility rate. LIDAR's execution involves the following steps:

**Step 1**: At startup, node IDs are arbitrarily assigned. Initial clustering of mobile nodes is performed using LID algorithm, chosen due to its simplicity, fast and inexpensive completion of clustering process.

**Step 2**: Throughout clustering maintenance phase, LID algorithm is invoked at the end of every HP, adjusting cluster formations according to current topology status; as discussed later, the duration of HP is dynamically adjusted according to the mobility rate of network nodes.

**Step 3**: During a period $P_{LIDAR} = k * HP$ (where the value of constant *k* is pre-defined), each mobile node *v* calculates the following weighted function value:

$$W_v = w_1 B_v - w_2 M_{v,P}, \; w_1 + w_2 = 1 \qquad (1)$$

where $B_v$ denotes the remaining battery life of node *v* and $M_{v,P}$ represents the mean mobility rate of node *v* within the latest $P_{LIDAR}$ period (in the following section, we describe how mobility rate is measured).

**Step 4**: $W_v$ values are unicasted by mobile nodes to their local CH (an alternative way would be to send $W_v$ values to a unique node, e.g. the node with the lowest ID in the network; however, such a scheme would cause excessive inter-cluster control traffic and create a bottleneck point).

**Step 5**: Having received $W_v$ values from their attached cluster members, CHs sort them in descending order and re-assign node IDs so that small IDs are assigned to nodes with larger $W_v$ values and large IDs to nodes with smaller $W_v$ values. Namely, lower IDs are assigned to nodes with sufficient power level and low mobility rate, thereby increasing their probability of being elected as CHs in the next algorithm's step.

**Step 6**: CHs send to their attached members their respective new_ID values.

**Step 7**: Mobile nodes update their ID values. Right after, re-clustering procedure is invoked (go back to Step 2).

LIDAR execution steps are illustrated in Figure 1. Table I presents how $W_v$ values are calculated, where the coefficients of equation (1) are set to $w_1 = 0.7$ and $w_2 = 0.3$:

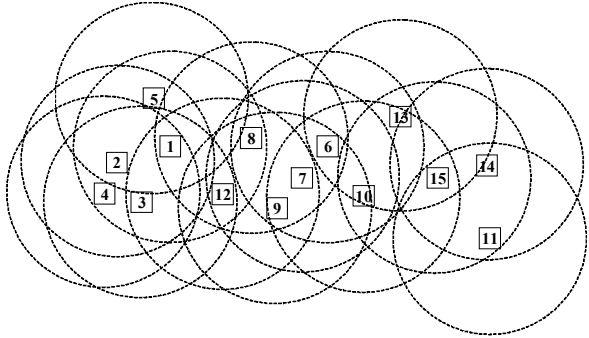

(a) Placement of mobile nodes on the plane after $k * HP$ time units (dashed circles indicate nodes transmission range)

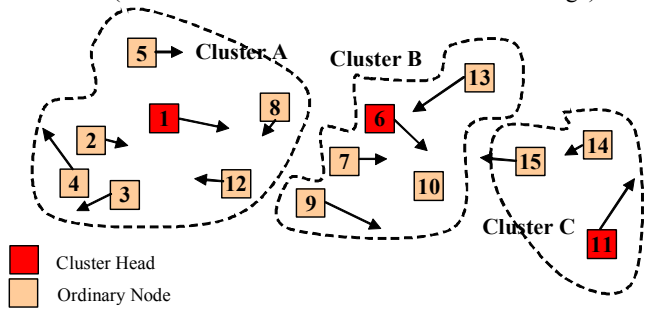

(b) Current clustering status of mobile nodes, based on lowest ID (arrows depict velocity and direction of nodes movement)

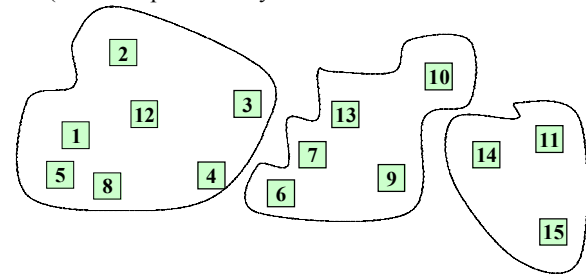

(c) Re-assignment of node IDs within individual clusters

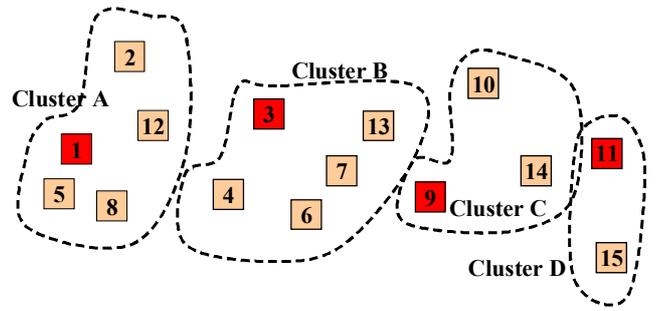

(d) Re-clustering of mobile nodes, based on lowest ID

Figure 1. Illustration of LIDAR execution steps.

TABLE I
CALCULATION OF $W_v$ VALUES AND NODE IDS RE-ASSIGNMENT IN LIDAR
(WHERE $w_1 = 0.7$ AND $w_2 = 0.3$).

|  | Node ID | $B_v$ | $M_{v,P}$ | $W_v$ | New Node ID |
|---|---|---|---|---|---|
| Cluster A | 1 | 2 | 4 | 0,2 | 12 |
|  | 2 | 7 | 1 | 4,6 | 1 |
|  | 3 | 4 | 3 | 1,9 | 8 |
|  | 4 | 6 | 4 | 3 | 5 |
|  | 5 | 7 | 2 | 4,3 | 2 |
|  | 8 | 6 | 1 | 3,9 | 3 |
|  | 12 | 6 | 2 | 3,6 | 4 |
| Cluster B | 6 | 3 | 3 | 1,2 | 13 |
|  | 7 | 7 | 2 | 4,3 | 7 |
|  | 9 | 8 | 4 | 4,4 | 6 |
|  | 10 | 6 | 0 | 4,2 | 9 |
|  | 13 | 7 | 4 | 3,7 | 10 |
| Cluster C | 11 | 3 | 4 | 0,9 | 15 |
|  | 14 | 6 | 1 | 3,9 | 11 |
|  | 15 | 6 | 2 | 3,6 | 14 |

### A. Mobility rate measurement

Most existing methods for estimating nodes mobility rate $M$ pose the requirement for GPS card with sufficient accuracy mounted on every mobile node (e.g. [2]). We propose an alternative method for measuring $M$ which relaxes mobile nodes from such requirement.

Each node $v$ measures its own mobility rate $M_v$ through contrasting the topology information it obtains during successive HPs. Mobile nodes maintain a short 'topology history table' (THT); THT rows comprise vectors representing the IDs of neighboring nodes, where each THT row refers to different HP. Calculated $M_v$ value actually represents the mean 'vector distance' among vectors recorded by $v$ during the latest $p$ HPs (where $p$ is a small integer in order to minimize memory requirement): $M_{v,t} = (1/p) \sum_{i=0}^{p-1} \left| \overline{THT_{t-iHP} - THT_{t-(i+1)HP}} \right|$, where $t$ denotes the current time.

Figure 2 illustrates how mobile node with ID = 1 moves on the plane; as a result of that movement (and the movement of other network nodes), its neighboring nodes (i.e. those within its transmission range) differ at the end of every HP. For this particular example, the 'neighborhood vectors' of node #1 at the end of four successive HPs: are $THT_1 = \{2, 3, 4, 5, 8, 12\}$, $THT_2 = \{2, 3, 5, 9, 12\}$, $THT_3 = \{2, 3, 5\}$, $THT_4 = \{3, 8, 12, 14\}$. Hence, the mobility rate of node #1 within this period of time is given by:

$$M_1 = (\overline{|THT_4 - THT_3|} + \overline{|THT_3 - THT_2|} + \overline{|THT_2 - THT_1|})/3 = (5+2+3)/3 = 3.33.$$

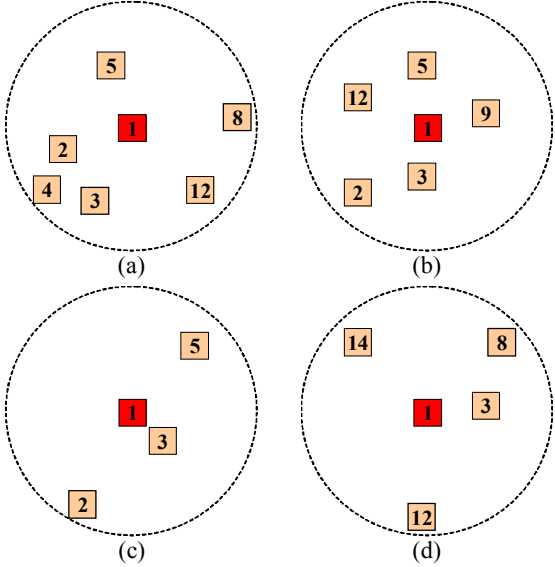

Figure 2. Neighboring nodes of node with ID = 1 at the end of four successive HPs.

### B. Minimizing cluster maintenance control overhead

Another objective of our algorithm is to minimize control traffic overhead during clustering maintenance phase, which highly depends on HP duration (i.e. frequency of broadcasting 'Hello' control packets). To achieve that, CHs measure the mean mobility rate of their attached cluster members $M_c$ (following the above described method) and accordingly adapt the 'Hello' broadcast period HP within their cluster. In particular, for highly mobile nodes ($M_c$ values), HP is shortened, i.e. message broadcasts are frequent enough to maintain consistent and accurate view of topology information. However, when nodes position on the plane does not considerably change over time relatively to their neighbors position (e.g. in conference sites or electronic classrooms), HP is lengthened, relaxing the MANET from unnecessary control message storms. It is guaranteed that HP duration always lies between two boundaries: $HP_{min} \leq HP \leq HP_{max}$; at startup, HP is globally set to $HP_{min}$. It is also stressed that potential HP synchronization problem among network nodes is prevented since *only* CHs are entitled to issue HP adaptation requests to their dominated nodes. In case of node migration to a neighboring cluster, its new CH informs the node about the HP of the local cluster.

The initial ideas behind our method for cluster maintenance overhead minimization have been described in [5].

## IV. SIMULATION RESULTS

The NS-2 simulator package [10] has been used to simulate the LIDAR algorithm and compare its performance against LID, HD and WCA algorithms. Our simulation tests attempt to compare the performance of these algorithms in terms of control traffic overhead and variance of MHs energy level.

We assume 50m MHs moving within a square terrain of 600m × 600m. At startup, MHs are randomly positioned on the plane. MHs move with speed 0 - 15m/s, on random direction. The 'hello period' duration is set to HP = 5ms for LID amd HD. The same HP value also applies at LIDAR's startup (during algorithm's execution phase, HP is adjusted according to MHs mobility pattern) and for cluster formation phase of WCA. Initial remaining battery time of MHs is randomly set between 20 and 100 units; energy is assumed to be linearly decreased for ordinary nodes, while for CHs it depends on the number of their attached cluster members. Each simulation run lasts 3 minutes; simulation results presented below have been averaged over 5 runs. Regarding the execution parameters of LIDAR, $W_v$ values are calculated for $w_1 = 0.7$ and $w_2 = 0.3$; MHs measure their mobility rate through contrasting the topology information they obtain during $P_{LIDAR} = 5$ HPs

Figure 3 illustrates the average number of control messages exchanged as a function of MHs average speed. In LID and HD algorithms, 'Hello' messages are periodically broadcasted during cluster maintenance phase; hence, their performance results coincide. In contrast, both WCA and LIDAR algorithms' performance depends on the average speed of MHs. However, LIDAR is shown to perform better in MANETs that exhibit low mobility behavior; in such environments, HP period is lengthened so as to avoid frequent exchange of control messages.

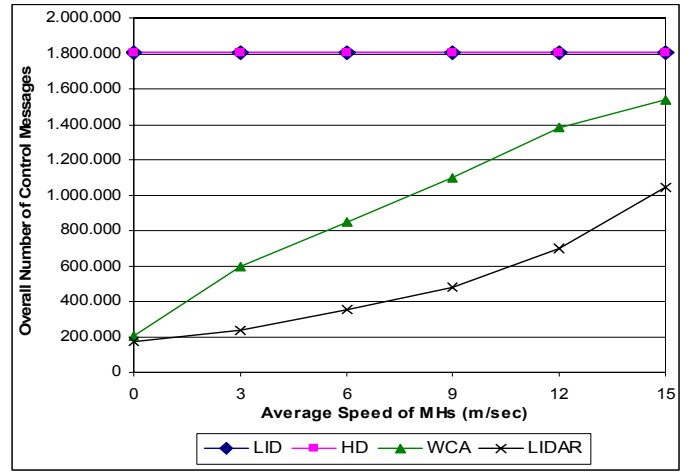

Figure 3. Overall number of control messages for 50 MHs.

Figure 4 illustrates the variance of power level among MANET's MHs. Large variance values indicate that specific nodes are engaged on CH role for long periods, hence, their energy level soon falls far below the average. This simulation test highlights the main limitation of LID algorithm: in LID, CHs election is biased in favor of nodes with low ID values;

these nodes are likely to serve as CHs for long time and their energy supply rapidly depletes. For static environments, WCA performs poorly: following the initial cluster formation, the lack of nodes movement prevents future re-clustering, hence CHs service time is prolonged and difference between the energy levels of CHs and ordinary nodes increases. However, higher mobility rates imply more frequent triggering of WCA re-clustering events, thereby decreasing variance values. LIDAR exhibits smaller variance of mobile nodes energy level: CHs give up their role even in static environments, when their battery resources are about to exhaust. Namely, CHs role is fairly shared among network nodes, achieving more uniform distribution of energy consumption.

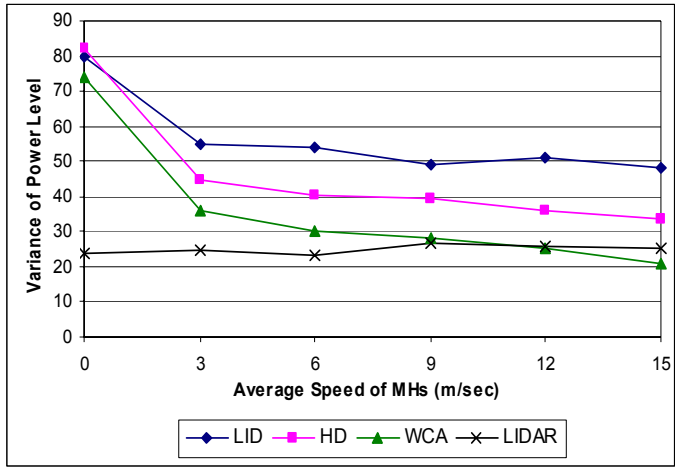

Figure 4. Variance of energy level among MHs for 50 MHs.

## V. CONCLUSIONS

We introduced an algorithm for efficient and energy-balanced clustering of mobile ad-hoc networks. Its contributions, compared to existing solutions, are summarized in the following: (a) clustering procedure is completed within two 'Hello' cycles; (b) both mobility and battery power metrics are taken into account in clustering process, so that suitable nodes are elected as CHs and energy consumption is uniformly distributed among network nodes; (c) for relatively static network topologies, control traffic volume is minimized; (d) fast packet forwarding and delivery is enabled, as clusters are pro-actively formed and topology information is available when actual user data exchange is required.

As a future extension, we intend to incorporate node degree metric within the calculated weight function, and also introduce a mobility prediction method in LIDAR. Finally, a variation of LIDAR algorithm where node IDs are received, sorted and re-assigned by a single (centralized) node will be implemented and compared in terms of cost against the distributed ID re-assignment method described in this article.